\documentclass[a4paper,11pt]{article}
\usepackage{jcappub}
\usepackage{lineno}

\usepackage{amsfonts}
\usepackage[utf8]{inputenc}
\usepackage{graphicx}
\usepackage{dcolumn}
\usepackage{bm}
\usepackage{aas_macros}
\usepackage{cleveref}
\crefname{equation}{equation}{equations}
\crefname{figure}{figure}{figures}

\newcommand{\mperiod}{\,.}

\newcommand{\D}{\mathrm{d}}

\usepackage{verbatim}

\title{The three phases of self-gravitating scalar field ground states}

\author[a]{Anthony E. Mirasola,}
\emailAdd{aem8@illinois.edu}
\affiliation[a]{%
    Department of Physics
    \\
    University of Illinois at Urbana-Champaign
    \\
    Urbana, IL 6180, USA
}

\author[b]{Nathan Musoke,}
\emailAdd{nathan.musoke@unh.edu}
\affiliation[b]{%
    Department of Physics \& Astronomy
    \\
    University of New Hampshire
    \\
    Durham, NH 03824, USA
}

\author[c,d]{Mark Neyrinck,}
\emailAdd{Mark.Neyrinck@hawaii.edu}
\affiliation[c]{%
    Institute for Astronomy\\
    University of Hawaii at Mānoa\\
    Honolulu, HI 96822, USA
}
\affiliation[d]{
     Blue Marble Space\\
     Seattle, WA 98104, USA
}

\author[b]{Chanda Prescod-Weinstein,}
\emailAdd{chanda.prescod-weinstein@unh.edu}

\author[e,f]{and J. Luna Zagorac}
\emailAdd{luna.zagorac@mcgill.ca}
\affiliation[e]{%
    Perimeter Institute for Theoretical Physics
    \\
    31 Caroline St. N. 
    \\
    Waterloo, ON N2L2Y5, Canada
}%
\affiliation[f]{%
    Department of Physics \& Trottier Space Institute
    \\
    McGill University
    \\
    Montr\'eal, QC H3A 2T8, Canada
}%

\date{\today}

\abstract{%
    It is generally assumed that scalar field dark matter halos would contain \emph{solitonic cores}---spherically symmetric ground state configurations---at their centers. This is especially interesting in the case of ultralight dark matter (ULDM), where the solitons sizes are on the order of galaxies. In this work, we show that the paradigm of a spherically symmetric soliton embedded in the center of each halo is not the only plausible structure formation scenario when there are multiple, interacting scalar fields. In particular, sufficiently strong repulsive interspecies interactions make the fields separate, creating the possibility of scenarios where there are distinct solitons. In such models, the ground state configuration can fall into a number of different phases that depend on the fields' relative densities, masses, and interaction strengths. This raises the possibility that the inner regions of ULDM halos are more complex and diverse than previously assumed.
}
                              
\arxivnumber{2410.02663}


\begin{document}

\maketitle

\section{Introduction}

\begin{figure}
    \centering
    \includegraphics[width=1.0\columnwidth]{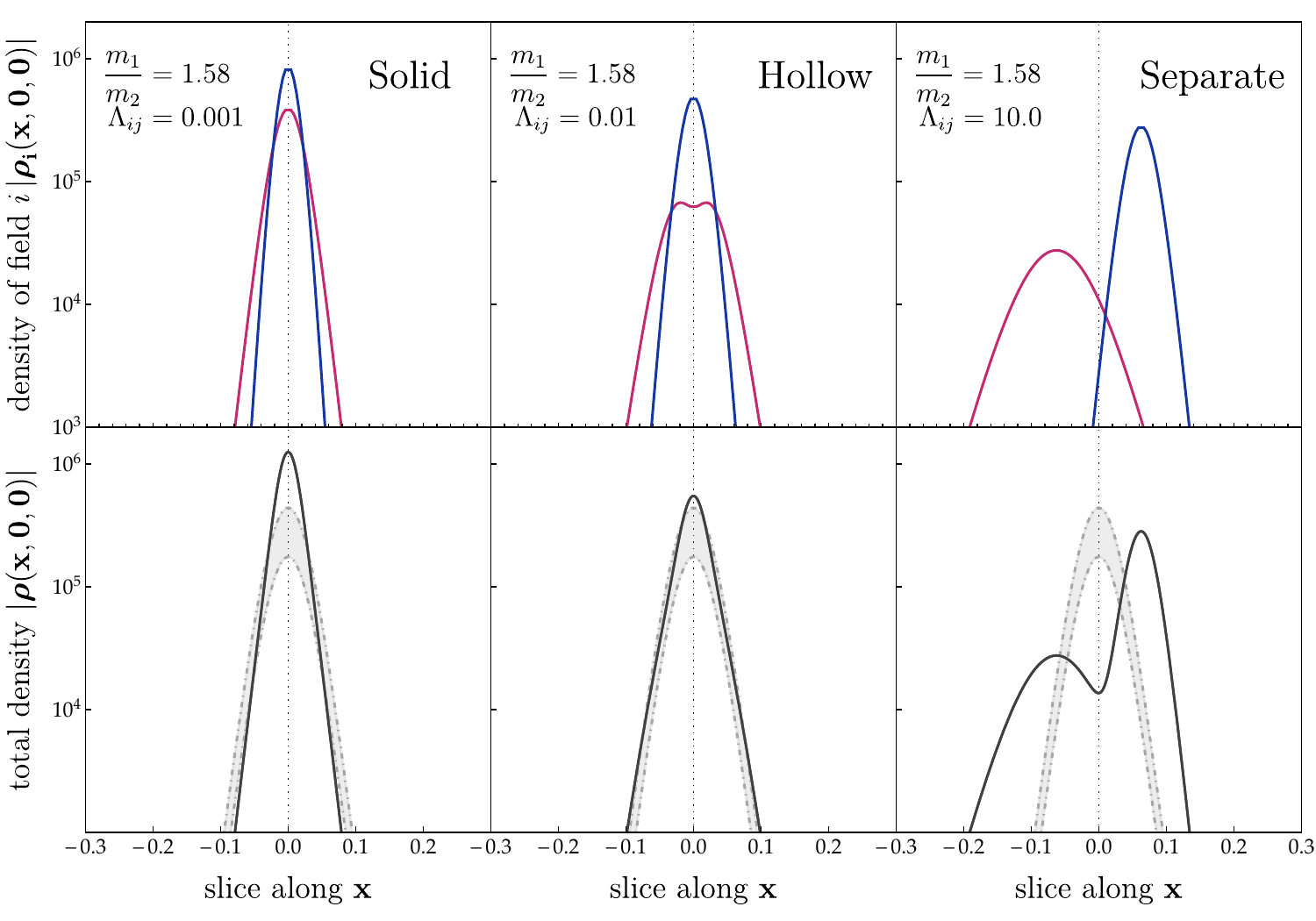}
    \caption{%
        Illustrations of the three classes of ground state configurations discussed in this paper. Each panel shows a slice through the center of a 3D numerical box. Each case has the same ratio of the two species' axion masses ($m_1$ and $m_2$, where $m_2 = 10^{-22} \, \mathrm{eV}$), but the interspecies interaction strength ($\Lambda_{ij}$, presented in code units described in Appendix A) increases from left to right. Upper panels show the densities of the individual fields; lower panels show the corresponding total density. For comparison, single-field solitons of equivalent total and particle masses are shown in light gray.  \textbf{The existence of three distinct classes of ground states and the resulting diversity in halo density shapes is the main result of this work. 
    }
    \label{fig:gs-types}
    }
\end{figure}

\begin{figure} 
    \centering
    \includegraphics[width=\columnwidth]{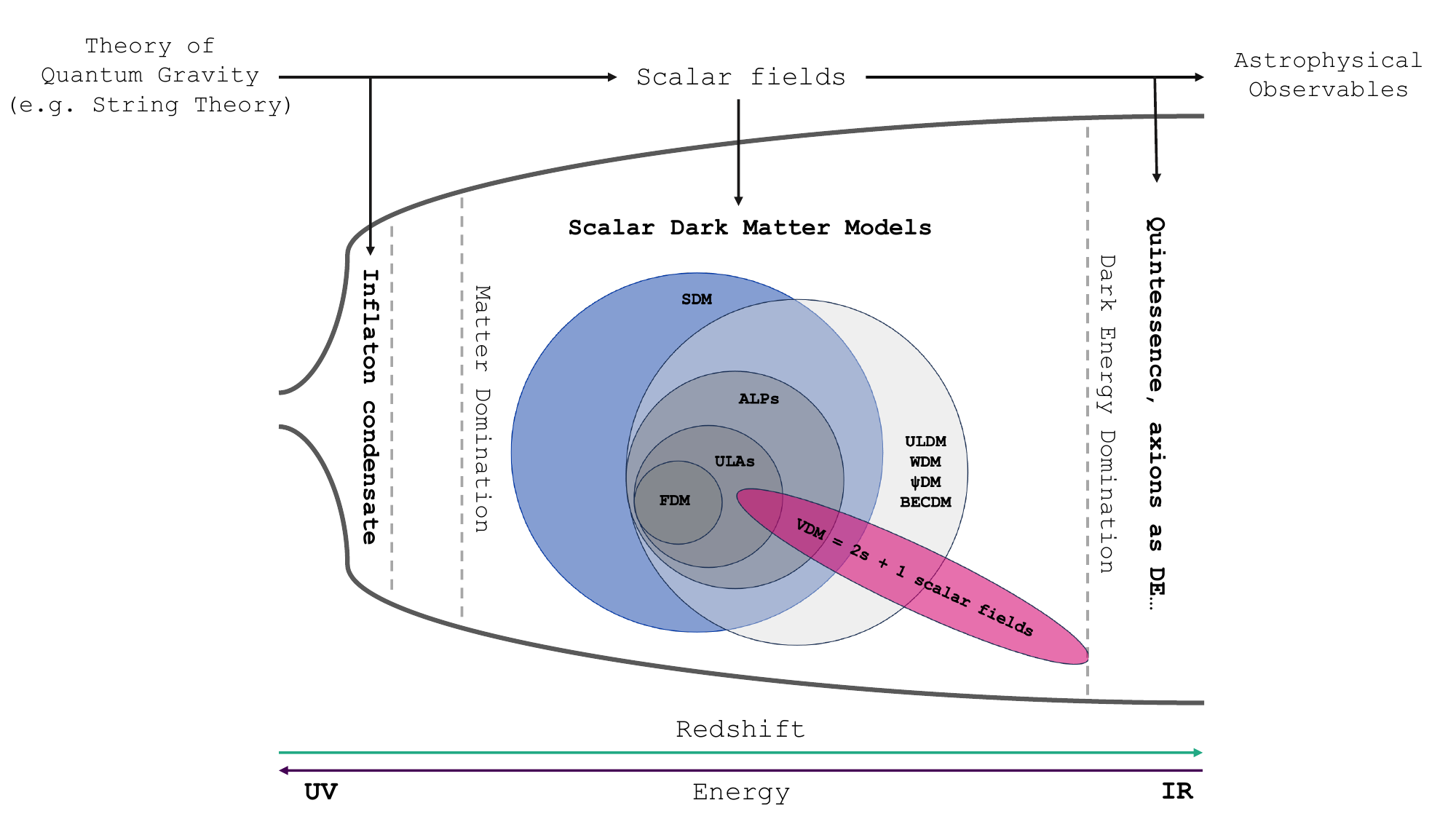}
    \caption{Scalar fields appear in many areas of cosmology, including as answers to the open questions of inflation, dark matter, and dark energy. We will particularly concern ourselves with their role as dark matter consisting of two ultralight axion-like particles (ULAs), though our results easily generalize to other applications. Considering how various scalar fields' phenomenologies impact astrophysical observables could help narrow down the parameter space of theories of quantum gravity.  
    }
    \label{fig:scalar-fields}
\end{figure}

Gravitating scalar fields are important components of theoretical models in cosmology: they arise naturally from certain overarching theories of quantum gravity and can have a variety of observational consequences in the Universe. In this paper, we discuss the ground states and equilibria of one family of real scalar fields. We are primarily interested in how these results can be applied to models of scalar dark matter (SDM) while also suggesting applications to models of inflation that include multiple scalar fields. The two applications are united in their reliance on the models' ground states, which can inform how structures collapse and grow in both regimes, directly relating to the focus of this work. Our main result is that there three distinct classes of ground states in such theories given the existence of repulsive non-gravitational interspecies interactions\footnote{{We not that what we call ``interspecies interactions" may sometimes be referred to as ``couplings" elsewhere in the literature.}} between the fields. Examples of each class are shown in Fig.~\ref{fig:gs-types}, and the roles gravitating scalar fields play in cosmology are outlined in Fig.~\ref{fig:scalar-fields}.  This is, to our knowledge, the first instance of non-spherically symmetric rotation-free ground states {applied in the context of the axiverse to cosmological scalar fields}.

Both in studies of inflation and SDM, multifield models are not new. Indeed, models of inflation that include multiple scalar fields have been studied in some depth starting from the 1980s ~\cite{Guth:1980zm, Linde:1981mu, Albrecht:1982wi, Martin:2013tda}.  On the SDM side, a well-known motivation comes as a consequence of string theory: the string axiverse generically predicts the existence of multiple ultralight axion scalar fields with nonzero self- and interspecies interactions \cite{Arvanitaki_2010, Arvanitaki_2011}.  We are therefore particularly interested in overlapping versions of ultralight dark matter (ULDM)\cite{Ferreira_2021}. There are several names for this class of models in the literature, and we take this opportunity to disambiguate our usage of the terminology and the assumptions behind each acronym in the following paragraphs. We also present the relationship between these terms graphically in the form of a Venn diagram embedded in Fig.~\ref{fig:scalar-fields}. 

As the names suggest, SDM can refer to any dark matter model made up of scalars, while ULDM simply requires that its constituent particles be ultralight in mass (often taken to mean $m \lesssim 10^0 \mathrm{eV}$ \cite{Ferreira_2021}). One feature of dark matter being ultralight is that its wave-like nature becomes consequential on the spatial scale set by the de Broglie wavelength, $\lambda_{\mathrm{dB}} \propto m^{-1}$; thus, such a dark matter model is also often referred to as wave dark matter\footnote{We use the acronym WDM for simplicity in Fig.~\ref{fig:scalar-fields}, but don't advocate for its wider use to avoid confusion with warm dark matter.} or $\psi$DM. Furthermore, if the constituent particles are also bosons (such as axion-like particles, or ALPs, such as would arise in the axiverse\footnote{These can also sometimes be called string axions to distinguish them from the QCD axion which solves the strong CP problem and stems specifically from the Peccei-Quinn mechanism \cite{1977PhRvL..38.1440P}.}), they will form Bose-Einstein condensates on the same scales. 

These Bose-Einstein Condensate Dark Matter (BECDM) models feature a Bose-Einstein condensate-like core often called a soliton. In single-field scenarios, the soliton corresponds to the ground state of the system and has a well known profile \cite{1992PhR...221..251L, 2014PhRvL.113z1302S, 2014NatPh..10..496S}, though the exact shape is modulated by the presence of self-interactions \cite{Chavanis2011}. The size of the soliton is inversely proportional to the mass of the constituent boson $m$; thus, different models admit solitons on different scales. Because a soliton is essentially the smallest collapsed structure that can be formed in a single-boson field, this means different models build or modify structure on different scales: for instance, stellar-sized solitons are sometimes referred to as Bose stars \cite{1992PhR...220..163J, Levkov:2018kau}.
For a particle mass of $m \sim 10^{-19} - 10^{-22}$, the de Broglie wavelength becomes evident on galactic scales ($\lambda_{\mathrm{dB}} \sim 1 \, \mathrm{kpc}$) \cite{Hu_2000}; this specific model is often referred to as fuzzy dark matter (FDM) and assumes the particle has no non-gravitational interactions. If such interactions are permitted, the model has been referred to as ultralight axions (ULAs).  Finally, while we began our discussion with scalar fields, there is also overlap with some models of vector dark matter (VDM); we specifically refer to cases with multiple species of massive particles with integer spin $s$, mathematically equivalent to $2s+1$ scalar fields \cite{Jain:2021pnk, Amin:2022pzv}. 

While our work presented here is generically applicable for many of the above mentioned models, we present results centered on parameter choices relevant to the ``fuzzy" regime, where ground states or solitons correspond in size to galaxy cores.  Most work on single-field ULDM, even when not analyzed as a Bose-Einstein condensate, assumes that each dark matter halo has a \emph{solitonic core} at its center. In ULDM, solitons condense quickly enough that most halos would contain a soliton~\cite{Levkov:2018kau, Eggemeier:2019jsu, Kirkpatrick:2020fwd, Chen:2020cef, Hertzberg:2020hsz, Kirkpatrick:2021wwz, Garcia:2023abs}. Indeed, the role solitons could play as a solution to the cusp-core problem contributed to ULDM rising in popularity as a dark matter candidate~\cite{Su_2011,Hui:2016ltb}. It is generally understood that solitons exist in halo cores and are roughly spherical, although recent work has clarified that there is some diversity in how they relate to underlying haloes {and baryonic structures in the galaxy}
\cite{2019PhRvD..99j3020B,2022MNRAS.511..943C,2022PhRvD.105b3512Y,2022MNRAS.511..943C,Kendall:2023kit,2023PhRvD.107h3513Z,Dave_2023,Dave_2024,Chakrabarti_2022}. Furthermore, while solitons are a feature of ULDM halos, it is important to emphasize that they are not a stationary structures and do exhibit oscillations,  fluctuations, and motions within the halo core (see e.g. \cite{Li_random_walk}); thus, it is important to emphasize that while understanding the basic soliton structures of a model are a crucial first step, their behavior in vacuum is not fully representative of their behavior in a realistic dark matter halo. 

Moving beyond single-field dark matter models, studies of ULDM have begun to consider how solitonic structures would look in multifield scenarios such as the axiverse. Changing the number of dark matter fields and allowing for non-gravitational interactions amongst them certainly impacts the ground states (and therefore the core profiles) of such structures~\cite{Luu:2018afg, Eby:2020eas, Guo:2020tla, 2021PhRvD.103j3501A, 2023PhRvD.107h3014G, Glennon:2023jsp, Luu_2023}. The rising levels of interest in such axiverse scenarios are supported in the recent literature \cite{2023PhRvD.107h3014G, Glennon:2023jsp, Huang_2023, Jain:2023ojg, 2024PhRvD.109d3507L}, including a particular emphasis on two-field scenarios and their core properties \cite{Luu:2023dmi,van_Dissel_2024}. Thoroughly understanding the predictions of the simplest multi-field scenario is crucial for building intuition for more complicated axiverse predictions. Indeed, comparing \textit{qualitative} differences in predicted ground state profiles with the latest data could help us constrain huge swathes of the axiverse parameter space, thus focusing our attention on the most astrophysically viable and interesting cases. Thus, it is valuable to understand the mapping between hypothetical interaction strengths and the ground states and the core profiles in which they ultimately result. A comparison between these results and the diversity of halos found in nature could constrain axiverse models, and even other non-axionic dark sectors \cite{2025arXiv250915299C}. 

These same results also have implications for inflation. In some inflationary models, the reheating epoch is analogous to structure formation in ULDM~\cite{Easther:2010mr, Musoke:2019ima}, going as far as the formation of inflaton halos and solitonic inflaton stars~\cite {Eggemeier:2019jsu}.
Interactions between multiple scalar inflaton fields would significantly change these collapsed states, impacting their dynamics during reheating and the consequent observational predictions~\cite{Hotinli:2017vhx}. Each of these applications will, of course, be highly model-dependent.  

In this work, we present a detailed exploration of the equilibria of two-species models with quartic interspecies interactions. {We restrict ourselves to understanding the qualitative features of a useful toy model, with a future goal of} surveying a parameter space of interaction strengths and masses consistent with constraints from the string theory axiverse \cite{Mehta:2021pwf,mehta_superradiance_exclusions}. We find that there is a wider variety of possible equilibria than previously known: in particular, sufficiently repulsive interactions can result in the separation of the two fields. This separation of the two fields mimics behavior that occurs in dual species Bose-Einstein condensates of ultracold atoms, which can have mixed (or ``miscible") and separated (or ``immiscible") phases  \cite{Ho_1996,Myatt_1997,Esry_1997,Timmermans_1998,Lee_2016,Jiang_2019,Gutierrez_2021}.
Even when numerically imposing the assumption of spherical symmetry on the system, sufficiently repulsive interactions imply the existence of ground states in which the maxima of the two species are not co-located, i.e., with a density maximum in a shell a nonzero radius from the center (see middle column in Fig.~\ref{fig:gs-types}).
Upon relaxing the assumption,  we find that ground states do not necessarily maintain spherical symmetry on their own and instead favor axisymmetric configurations (see right column in Fig.~\ref{fig:gs-types}).
This picture would become increasingly complex with more than two species of scalar particles; nevertheless, this asphericity presents new avenues for constraining this interesting but enormous parameter space.

\section{Analytically Evaluating Multispecies ULDM for Phase Transitions}

\subsection{The System of Equations and Relevant Energies}

We begin by concretizing the ULDM model we consider here. In general, multispecies ULDM is governed by a system of coupled Gross-Pitaevskii-Poisson (GPP) equations:
\begin{gather}
    \label{eq:gpp}
    i \hbar \frac{\partial \psi_j}{\partial t}
    =
    - \frac{\hbar^2}{2 m_j} \nabla^2 \psi_j
    + m_j \Phi \psi_j
    + \frac{\hbar^3}{2 m_j^2} \lambda_{jj} |\psi_j|^2 \psi_j
    + \frac{\hbar^3}{4 m_j} \sum_{k \ne j} \frac{\lambda_{jk}}{m_k}  |\psi_k|^2 \psi_j
    \\
    \label{eq:poisson}
    \nabla^2 \Phi
    =
    4 \pi G \sum_j m_j |\psi_j|^2
\end{gather}
where $\Phi$ is the Newtonian gravitational potential, and the indices $j,k$ run over the $N$ fields considered, each with particle mass $m_j$.
The self- and interspecies interactions are parameterized by a symmetric matrix $\lambda_{jk}$ with positive (negative) values corresponding to repulsive (attractive) interactions. The GPP equations arise from the Newtonian limit of the ULDM Lagrangian \cite{Glennon:2023jsp}.

The mass density of each species is
\begin{equation}
    \rho_j(\mathbf{x}, t)
    =
    m_j \left|\psi_j(\mathbf{x}, t)\right|^2
   ,
\end{equation}
and the total mass in each field is independently conserved.

The conserved energy of \cref{eq:gpp,eq:poisson} can be written as
\begin{align}
    \label{eq:energy}
    E
    &=
    E_{\text{grav}} + E_{\text{KQ}} + \sum_j E_{\text{self}}^{j} + \sum_j \sum_{k > j} E_{\text{int}}^{j, k} \mperiod
\end{align}
In order, there four energies refer to the gravitational potential energy, the kinetic energy, the energy of self-interactions, and the energy of interspecies interactions. We further define each type below, and they will be crucial to the arguments we make in subsequent sections. 

The gravitational potential energy is
\begin{equation}
    \label{eq:e_grav}
    E_{\text{grav}}
    =
    \int \D x^3 \frac{1}{2} \Phi \sum_j m_j |\psi_j|^2
    \mperiod
\end{equation}
The total kinetic energy consists of a classical term (arising from bulk motion) and quantum term.\footnote{Note that the ``quantum'' kinetic energy does not have a quantum origin, but simply refers to the kinetic energy that does not correspond to any bulk motion~\cite{Niemeyer:2019aqm}.}
Throughout this work we only consider equilibria, so the classical kinetic energy is zero and the total kinetic energy will be equal to the quantum kinetic energy:
\begin{equation}
    \label{eq:e_k}
    E_{\text{KQ}}
    =\frac{\hbar^2}{2}
    \int \D x^3  \sum_j \frac{1}{m_j}\left| \nabla \psi_j \right|^2
\end{equation}

The energies due to self-interactions and interfield interactions 
respectively are 
\begin{align}
    E_{\text{self}}^{j}
    &=
    \frac{\hbar^3 \lambda_{jj}}{2 m_j ^2}\int \D x^3 |\psi_j|^4
    \\
    E_{\text{int}}^{j,k}
    &=
    \frac{\hbar^3 \lambda_{jk}}{2 m_j m_k}\int \D x^3  |\psi_k|^2 |\psi_j|^2
    \mperiod
    \label{eq:Eint}
\end{align}

In our investigations of the conditions of the system's immiscibility---or separation of ground states, as presented in the rightmost panel of Fig.~\ref{fig:gs-types}---we will make particular use of the energy definitions above.  

\subsection{Parameter choices and existing constraints}

Before diving into the results of this work, we briefly comment on the parameters we center our exploration around in the context of existing constraints. Our results are present in scalable code units, discussed in more detail in Appendix~\ref{app:code-units}.

Throughout this work, we center our exploration on a particle mass of $m_1 = m_0 \equiv 10^{-22} \, \mathrm{eV/c^2}$ in one field, with the mass of the other parametrized as a ratio $m_2/m_1$. In the FDM scenario with no self-interactions, this region is heavily constrained \cite{Ferreira_2021}\footnote{A useful summary of single-field limits can also be found on the website of Dr. Keir Rodgers at  \url{https://keirkwame.github.io/DM_limits/}.}. Given that it is the canonically ``fuzzy" regime, we begin our investigation in the same place, with a view to eventually evaluating which mass constraints may hold in the interacting multifield regime. 

The effect of self-interactions on ULAs has been explored elsewhere in the literature; they change the profile of ground state solitons, but do not fundamentally change their shape~\cite{Glennon:2023jsp,Guo:2020tla,Eby:2020eas,Luu:2023dmi}. Nevertheless, adding self-interactions to the single field model imposes certain limits: Ref.~\cite{2016arXiv160306580F} derived limits on mass and self-interaction strengths below the fuzzy regime by comparing pressure from self-interaction to the quantum pressure, while Ref.\ \cite{Cembranos2018} derives a constraint from the CMB power spectrum,
\begin{equation}
    \log \lambda < -91.86 + 4\log\left(\frac{m}{m_0}\right),
\end{equation}
which in our normalized units works out to $\Lambda \lesssim 3 \Lambda_0$. 

Interspecies interactions haven't been subject to many astrophysically-motivated dark matter constraints thus far. Certain models of string theory can provide guidance: for instance, in  Ref.~\cite{Mehta:2021pwf} the authors use black hole superradiance to predict interspecies interactions are not expected to exceed the self-interaction strengths~\cite{Mehta:2021pwf}. Given they are the most unconstrained feature of the model we explore, we were motivated to focus on exploration on how they affect potential observables. Consequently, we chose to put aside the effects of self-interactions by setting $\lambda_{ii} = 0$ and only varying $\lambda_{ij}$. 
This allowed us to explore the effects of growing interspecies interactions without adding to an already high-dimensional space of variables; the detailed exploration of the effects of self-interactions we leave for future work. {Furthermore, while we have been endeavored to be specific and clear in this text, any remaining ambiguity in the following sections when using the just the term ``interaction" should be understood to mean ``interspecies interaction."}

\subsection{Analytic evidence for phase transitions}
In both the single and multi-field scenarios, the GPP equations (\ref{eq:gpp}-\ref{eq:poisson}) admit no closed form solutions.
Numerical solutions for the ground state profile of a single self-gravitating scalar field are well studied \cite{2016PDU....12...50P, 2018JCAP...10..027E, 2020PhRvD.102h3518S, 2021PhRvD.104h3532G}.
They are static spherically symmetric states with balanced gravitational and ``quantum''/gradient forces, and are generally called \emph{solitons} in the literature.
Though exact closed form solutions do not exist, there are a number of well-studied approximations to the ground state soliton solution that can be useful for gaining physical insight and intuition.
For a single self-gravitating field, the most robust ground state approximation is a power-law profile which has a high agreement to simulations \cite{2014PhRvL.113z1302S}, 
\begin{equation}
    \label{eq:soliton_profile}
    \rho(r)
    =
    \frac{\rho_s}{{\left(1 + {(r/r_s)}^2\right)}^8},
\end{equation}
where $\rho_s$ is the central density and the scale radius $r_s$ is
\begin{equation}\label{eq:sol-rad}
    r_s \simeq 0.335 \, \frac{10^9 M_{\odot}}{M}\left(\frac{10^{-22} \mathrm{eV}}{m}\right)^2 \, \mathrm{kpc}
    \mperiod
\end{equation}
The total mass of this profile is
\begin{equation}
    M
    = 2.2 \times 10^8\left(\frac{m}{10^{-22} \mathrm{eV}}\right)^{-2}\left(\frac{r_s}{\mathrm{kpc}}\right)^{-1} M_{\odot}
    \mperiod
\end{equation}
Note that the radius of a soliton decreases with both the mass $m$ of its constituent particles and its total mass $M$. The above equations refer to the minimal FDM model; the addition of self-interactions affects the possible types of equilibria and imply maximum soliton masses~\cite{Chavanis:2022fvh}. 

The profile of non-isolated ground states is expected to be more complicated than the single-field soliton approximated by  \cref{eq:soliton_profile}, which is conveniently universal when scaled by particle and soliton mass, $m$ and $M$. As with single-field modes, the GPP equations (\ref{eq:gpp}-\ref{eq:poisson}) governing multifield scenarios admit no closed form solutions. Furthermore, each field $\psi_j$ depends on the other fields $\psi_{k \ne j}$, already suggesting that universality of ground state profiles may be broken. We begin to build intuition about these states with energy minimization arguments using an analytic approximation for each field's profile. The power law form of the above approximation makes using it for calculating derived quantities such as energies unwieldy, thereby losing the advantage of the analytic approach. For our purposes, a Gaussian profile is sufficient for approximating isolated solitons, and is useful for producing clean analytic estimates. 

Thus, before solving our system numerically, we present derivations of exact expressions for the energies under a spherically symmetric Gaussian ansatz for the ground state profiles. For single-species ULDM solitons, this ansatz has close agreement with numerical solutions \cite{Chavanis2011,2021PhRvD.104h3532G}. We assume each field has the profile defined by its mass $M_i$ and length scale $R_i$, with centres of the profiles separated by a distance $d$: 
\begin{equation}
    \rho_i(\mathbf r) = \frac{M_i}{\pi^{3/2} R_i ^3}\exp\left[-r^2/R_i ^2\right].
\end{equation}
With this ansatz, we can find analytic expressions for the quantum kinetic, gravitational, and interaction energies. As mentioned above, the quantum kinetic energy does not depend on the separation distance $d$, and therefore plays no role in determining which phase is preferred. The interaction energy can be calculated from Eq. (\ref{eq:Eint}),
\begin{align}
    E_\mathrm{int}^{12}(d)
    &= \frac{\hbar^3\lambda_{12}}{4\mu^4} \frac{M_1 M_2}{\pi^{3/2} (R_1 ^2 +R_2 ^2)^{3/2}} \exp\left[-\frac{d^2}{R_1 ^2 +R_2^2}\right],
\end{align}
where $\mu^2=m_1 ^2 m_2 ^2/(m_1 ^2 + m_2^2)$. The gravitational energy has a self-gravitational component and interspecies gravitational component. The former is independent of $d$ and is easily calculated as
\begin{equation}
    E_{\mathrm{grav}}^{i}=\int d^3 \mathbf r \rho_i(\mathbf r) \Phi_i (\mathbf r) = \frac{-1}{\sqrt{2\pi}} \frac{GM_i ^2}{R_i},
\end{equation}
where $\Phi_i$ denotes the gravitational potential generated by species $i$.

To calculate the interspecies gravitational energy $E_\mathrm{grav}^{12}$, we consider two cases. First, when $d=0$ we can calculate it exactly in the Gaussian ansatz,
\begin{align}
     E_\mathrm{grav}^{12}(d=0)
     &= \int d^3 \mathbf r \left(\rho_1 \Phi_2 + \rho_2 \Phi_1\right)
     \\
     &= \frac{2 G M_1 M_2}{\sqrt{\pi}}\left[\frac{1}{\sqrt{R_1 ^2+R_2 ^2}}- \frac{R_{1} ^2}{(R_1 ^2 + R_2 ^2)^{\frac{3}{2}}} - \frac{R_{1} ^2}{(R_1 ^2 + R_2 ^2)^{\frac{3}{2}}}\right].
\end{align}
A simplification occurs when the two species have equal total mass $M$, particle mass $m$, and radii $R$. In this case, the total gravitational energy evaluates to
\begin{equation}
    E_\mathrm{grav}(d=0) = \frac{-4}{\sqrt{2\pi}} \frac{GM ^2}{R}. 
\end{equation}
This agrees with the total gravitational energy of a single-species soliton in the Gaussian ansatz. Second, at finite separation distance $d>0$, the interspecies gravitational energy $E_\mathrm{grav}^{12}$ must be evaluated numerically. However, it can be approximated at large separation distances $d\gg R_1,R_2$ by noting that in this case the solitons become effectively point masses such that
\begin{equation}
    E_\mathrm{grav}^{12}(d\gg R_i) = \frac{-G M_1 M_2}{d}.
\end{equation}

The gravitational energy $E_\mathrm{grav}$ and the energy $E_\mathrm{int}$ due to repulsive interactions are in competition with each other: $E_\mathrm{grav}$ is minimized at $d=0$, while $E_\mathrm{int}$ is minimized at $d=\infty$. This suggests that there is a critical interaction strength at which the energy minimizing separation distance changes from $d=0$ to $d>0$. We can estimate the critical interaction strength $\lambda_{12} ^*$ by comparing the total energy at $d=0$ to the total energy at $d=\infty$,
    $\Delta E = E(d=0) - E(d=\infty)$.
Since the quantum kinetic energy and self-gravitational energies are independent of $d$, and both interspecies gravitational energy and interaction energy go to zero as $d$ becomes infinite, this simplifies to
\begin{equation}
    \Delta E = E_\mathrm{int}(d=0)+ E_\mathrm{grav} ^{12} (d=0).
\end{equation}
When $\Delta E$ is positive, the system has lower total energy at infinite separation distance than at zero separation distance, indicating that the system is not in the nested miscible phase, meaning that the two solitons will have started to separate.
Thus, solving $\Delta E = 0$ gives an estimate for the critical interaction strength. When the two species have equal total mass and particle mass, this results in
\begin{equation}
    \lambda_{12} ^* \approx \frac{8\pi G R^2 m^4}{\hbar^3}.
    \label{eq:lcrit}
\end{equation}
The radius $R$ can be determined by variationally minimizing the energy of the Gaussian ansatz. Fixing to zero separation distance, this results in
\begin{equation}\begin{aligned}
    R &= \frac{3 \sqrt{\pi} \hbar^2}{4 \sqrt{2} G M m^2} \left(1+ \sqrt{1 + \frac{G M^2 m^2 \lambda_{12}}{6 \pi^2 \hbar}}\right).
\end{aligned}\end{equation}
Together with Eq.\ (\ref{eq:lcrit}), this allows us to estimate the critical interaction strength for a given total mass. For two fields of particle mass $m=m_0= 10^{-22}$ eV, and soliton mass $M=50 \mathcal{M} \approx 3.5\times10^9 {\left(10^{-23} \mathrm{eV} / m_0\right)}^{\frac{3}{2}}M_\odot$, this results in an estimated critical interaction strength of $\lambda_{12} ^* = 0.09 \Lambda_{12}$, where the code unit $\Lambda_{12}$ is defined in the Appendix. This closely approximates the critical interaction strength we observe in numerical simulations presented below, where we find $\lambda_{12} ^* = 0.1 \Lambda_{12}$.

\section{Numerical results and phase diagram}\label{sec:numerical-results}

Having established an expectation for a critical interspecies interaction value  $\lambda_{12}^*$ where a phase transition will occur, we now relax our Gaussian ansatz to find more realistic ground state profiles. Previous work suggests a starting point: in the absence of interspecies interactions, multi-species equilibria are essentially multiple coincident single-field solitons, scaled to account for a background gravitational field ~\cite{2023PhRvD.107h3014G, Glennon:2023jsp,Luu_2023, Huang_2023,van_Dissel_2024}. The overall ground states display spherical symmetry and coincident centres of mass for each field's soliton. The same follows from our energy argument above: when $\lambda_{ij}=0$, the only interspecies feedback is through the gravitational field and is strictly attractive.  Although the exact dependence of these solutions each parameter choice is nontrivial, the overall trend of $r_s \propto m_j^{-2} M_j^{-1}$ from the single-field case is preserved \cite{2014PhRvL.113z1302S}. In keeping with previous literature, we will call such configurations \emph{nested}~\cite{Luu:2023dmi}.

Let us now consider cases with $\lambda_{ij}>0$, thus introducing a competing repulsive interspecies feedback, and find the ground states of such systems numerically. We use the \texttt{nSPIRal} Mathematica module to find numerical solutions with multiple species and self-interactions.
\texttt{nSPIRal} is an extension of the code used in Ref.~\cite{Zagorac:2021qxq} to evolve the spherically symmetric version of the Schr\"{o}dinger-Poisson system in time. In this work, we use \texttt{nSPIRal} to evolve the spherically symmetric equations of motion for two fields forward in imaginary time to arrive at the overall system's ground state. Unlike the previous version, our version of \texttt{nSPIRal} allows for multiple axion fields, self- and interfield interactions, and the choice of real or imaginary time; see Appendix \ref{app:numerics} for more details.

We find imaginary time evolution to be a reasonable approach if we assume there is a spectrum of eigenstates that describes the system evolved by equations~\ref{eq:gpp} and~\ref{eq:poisson}. Any initial profile for each wavefunction is therefore some superposition of these eigenstates. Furthermore, each eigenstate has an associated eigenenergy, with the ground state having the lowest energy. Ignoring the backreaction on the gravitational potential for a moment, each wavefunction $\psi_j$ will evolve as
\begin{align}
  \psi_j(t) = \sum_{i=1}^{N} c^i_j \exp \Biggl(-i E_j^i t\Biggr) \phi_j^i \,, 
\end{align}\label{eq:lin-theory}
where $\phi_j^i$ is an eigenstate, $E_j^i$ its associated eigenenergy, and $c_j^i$ a complex expansion coefficient. Now, evolving in imaginary time $t \rightarrow i t$ causes the exponential to decay at a rate set by $E_j^i$. Given the ground state has the largest absolute value of its energy $|E_0|$, excited states will exponentially decay first, leaving us with the ground state profile.\footnote{This does not evolve the system in time, so it need not be unitary.  We also re-scale the mass of the system at each imaginary time step to ensure it is conserved in each field. More details of the calculation are available in the Appendix.} Thus, we arrive at the ground state of the system without \textit{a priori} knowing the full spectrum of our system of equations. 

For each simulation, we must first specify initial conditions for the $\psi_j$ soliton profiles. In principle, the utility of the imaginary time solver should not keenly depend on the initial conditions, with the understanding that 1) the true ground state is present in the eigenstate makeup of the guess and 2) the numerical parameter requirements for the solver to reach the ground state, such as total evolution time, could be impacted. We initialize using the multifield soliton profiles described in \cite{Glennon:2023jsp}, thus guaranteeing a large stake of the ground state at least in the $\lambda_{ij} \rightarrow 0$ limit. 

\begin{figure}
    \centering
    \includegraphics{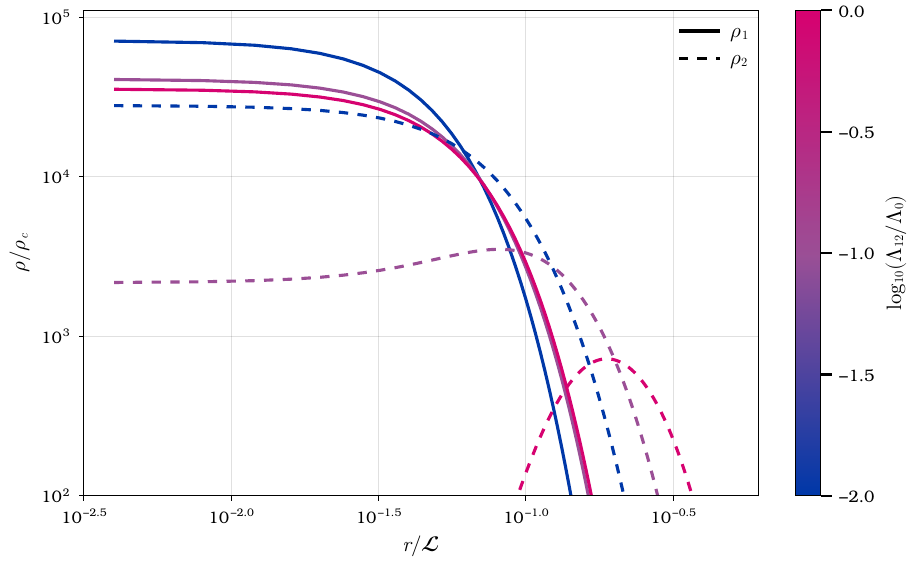}
    \caption{%
        Density profiles of numerically calculated spherically symmetric ground states.
        The solid lines are densities of $\psi_1$ and the dashed lines are densities of $\psi_2$.
        The color scale indicates the interaction strength $\lambda_{12}$; large positive values are repulsive.
        In each case, $M_1 = M_2 = 50 \mathcal{M}$, $m_1 = m_0$ and $m_2 = 10^{-0.1} m_0$.
        When the interaction strength is small, solutions are nested, similar to those presented in \cref{fig:profile_nested} and discussed elsewhere in the literature.
        The less massive field $\psi_2$ has a larger characteristic radius, as expected from \cref{eq:soliton_profile}.
        When interactions are strongly repulsive, we find hollow solutions, in which $\psi_2$ has a local minimum at $r=0$.
    }
    \label{fig:profile_hollow}
\end{figure}

We have also found that much of our parameter space is relatively insensitive to changes in our numerical parameters and even exact initial profile, with one important exception: when $m_1 = m_2$ and $M_1 = M_2$. The imaginary time evolution preserves the symmetry $\psi_1(r) = \psi_2(r)$, and only nested soliton-like solutions are found for any value of $\lambda_{ij}$. The two fields become very nearly degenerate (or exactly degenerate in the case of $m_1 = m_2$ and $M_1 = M_2$). When initializing  instead with some small perturbation to each profile such that  $\psi_1(r) \ne \psi_2(r)$ instead, we find solutions that break this symmetry. Removing the symmetry in the initial conditions leads to a set of solutions that are still nested, but no longer have coincident maxima; rather, one has field a density maximum that is set apart from the center in a shell with a nonzero radius from the center. We call this type of solution \textit{nested hollow}, and the type described above \textit{nested solid} for disambiguation. The transition from solid to hollow profiles with increasing $\lambda_{ij}$ is illustrated in Fig.~\ref{fig:profile_hollow}.  

Note that this nested hollow phase was not predicted by our analytic arguments from the previous section, which focused on changing distance between centres of mass of each field's soliton. 

\begin{figure}[ht]
    \centering
    \includegraphics[width=0.9\textwidth]{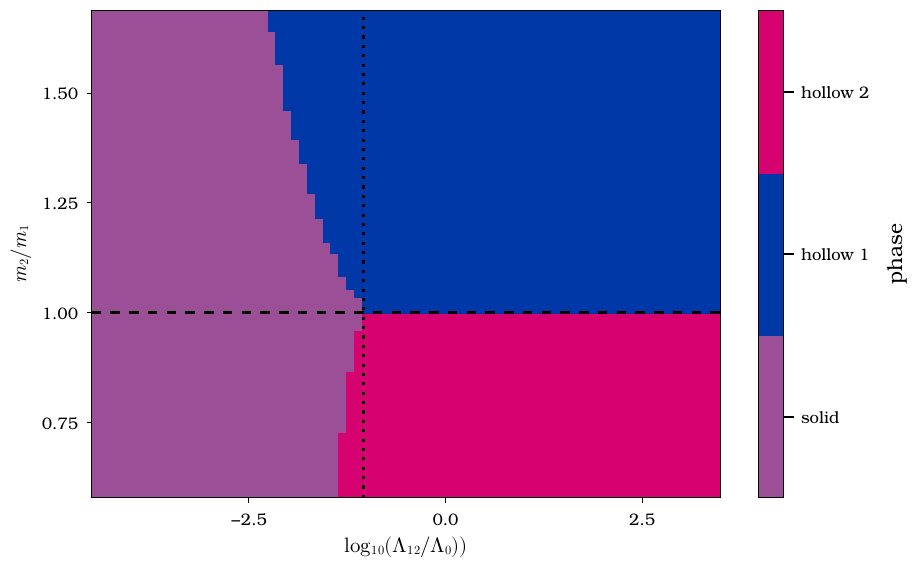}
    \includegraphics[width=0.9\textwidth]{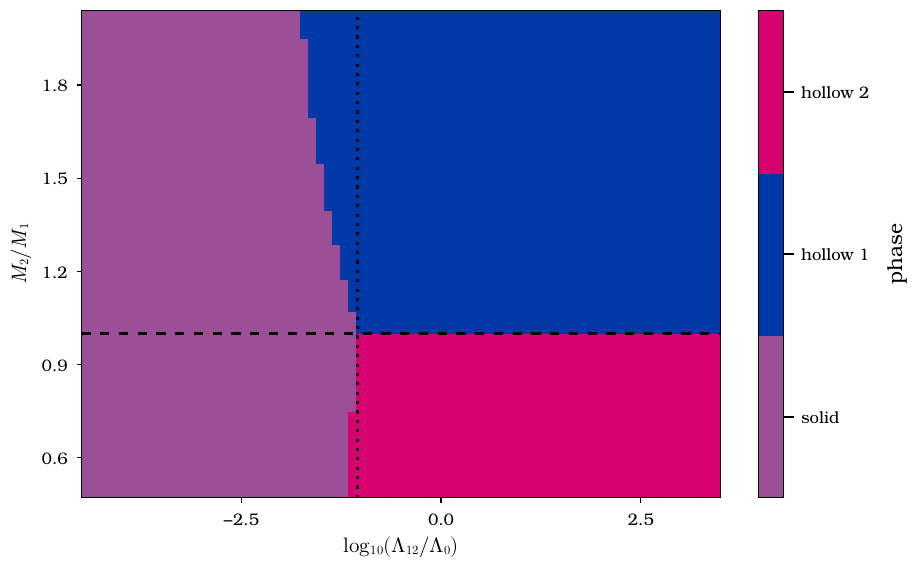}
    \caption{%
        Classification of spherically symmetric solutions found by relaxation in imaginary time, as a function of interaction strength $\Lambda_{12}$ and particle mass ratio $m_2/m_1$ (\textbf{top}) and total mass ratio $M_2/M_1$ (\textbf{bottom}).
        Points are classified as ``solid'' if both fields have a local maximum at $r = r_{\text{min}}$, or ``hollow 1'' (``hollow 2'') if $\psi_1$ ($\psi_2$) has local minimum at $r = r_{\text{min}}$.
        The dotted line shows the analytic prediction $\Lambda_{12} = 0.09$ for the transition from solid to hollow states.
        The dashed line shows the expected separation between ``hollow 1`` and ``hollow 2`` states at $r_{s, 1} = r_{s, 2}$.
        In the \textbf{top} panel, all points have $M_1 = M_2 = 50 \mathcal{M}$ and $m_1 = m_0$; in the \textbf{bottom}, $M_1 = 50 \mathcal{M}$ and $m_1 = m_2 = m_0$.
    }
    \label{fig:phase-1d-lambda}
\end{figure}

{However, readers familiar condensed matter literature might have expected the hollow phase by drawing a parallel to the `core-shell' \cite{PhysRevLett.80.1804} or `ball and shell` \cite{2008PhRvA..78e3613R} ground state found in an immiscible or partially miscible two-component BEC in a trap. To make the parallel explicit, the profiles of the two atomic species in the ``weak separation" regime (as shown in e.g. Fig.~1 of Ref.~\cite{2002PhRvA..66a3612B}) have the same structure as our nested hollow solutions. 
}
Here, even when keeping centers of mass coincident by construction of our numerical solver, the two fields' density peaks become no longer co-incident. We construct a phase diagram of these two nested cases by testing a range of parameters $\lambda_{ij}$, $m_1 / m_2$, and $M_1 / M_2$. In the region where the fields are nearly degenerate, we verify that we have reached the true ground state by running several choices of initial conditions and choosing the one that minimizes the energy of the system. The results are shown in Fig.~\ref{fig:phase-1d-lambda}. We center our investigation on some particle mass $m_0$ and mass in each of the fields of $50 \mathcal{M}$ for specificity and because we know this setup is well-resolved by our code. If we assume an FDM-like scenario, a particle mass of $m_0 = 10^{-22} \, \mathrm{eV/c^2}$ yields a mass in the field of approximately $10^8 \, M_{\odot}$, thus also presenting an interesting parameter space. 

Using our imaginary time solver, we have now identified two qualitatively different spherically symmetric phases and united them under the umbrella term nested (alternatively, mixed or miscible). In addition, the analytic arguments presented in the last section support a third: an unmixed or immiscible phase we call \textit{separate}, where the centers of mass of the two scalar field solitons are separated by a distance $d$. Such a configuration breaks spherical symmetry, and is therefore beyond the reach of our 1D imaginary time solver; however, we can use energy arguments to calculate the transition parameters instead. 
Recall that the gravitational potential energy consists of contributions from the self-gravitating configurations of each scalar field and a two-body term. The two-body gravitational energy is minimized when $d = 0$, while the interaction energy is minimized as $d\rightarrow\infty$. The quantum kinetic energy is agnostic with respect to the two fields' centers of mass, but is minimized when both density profiles are solitonic; the self-gravitating terms behave similarly. Therefore, a \textit{nested solid} state arises when gravitational and quantum kinetic energies are minimized at the expense of the interaction energy; a \textit{nested hollow} state arises when the interaction energy is lowered at the expense of the quantum energy of one of the states (whichever is ``hollow"); and the \textit{separate} state arises when the interaction and quantum energies are minimized at $d > 0$ at the expense of the two-body gravitational energy.

Thus, we expect the transition from miscibility to immiscibility will occur approximately when the energy budget of the quantum, interaction, and self-gravity energies is on the order of the 2-body gravitational energy
\begin{align}\label{eq:energy-condition}
    E_{KQ} + E_{\mathrm{int}} + \sum |E_{\mathrm{grav}}^{i}| \sim |E_{\mathrm{grav}}^{12}|.
\end{align}
Because the dependence of the energies on wavefunction profiles and their overlap cannot be fully disentangled from their dependence on the distance between the centers of mass, this argument cannot be easily leveraged to make analytic predictions on phase transitions. However, we include it here as useful intuition for why the three phases arise.

The existence of this separate phase raises the possibility that our \textit{nested hollow} solutions are artifacts of the 1D solver used to find them.
We verified that the hollow phase is \textit{not} just a numerical artifact of the 1D imaginary time solver by putting a range of corresponding wavefunction profiles on a 3D grid using \texttt{UltraDark.jl} \cite{Musoke:2024umh}.
For each case, we compared the energy of ground state profiles generated in 1D with the energy of two solitonic profiles of appropriate masses at a range of separations, allowing us to check if there are obvious cases in which a separate' configuration has lower energy than a `hollow nested' configuration.  Three examples of energy-minimizing configurations verified in 3D are illustrated in Fig.~\ref{fig:gs-types}. At relatively large interspecies interaction strengths, there are separate configurations that have lower energy than the corresponding nested hollow configurations.
This is how we arrived at the profile shown in the rightmost panel of Fig.~\ref{fig:gs-types}. 
We also observed a general trend that the energy-minimizing distance $d$ grows with increasing $\lambda_{ij}$.
However, at lower interaction strengths, we were did not find separate configurations that minimized the energy, suggesting that many points are correctly classified as ``hollow 1" or ``hollow 2" in Fig.~\ref{fig:phase-1d-lambda}. 

Taken together, these findings suggest an additional region indicating a ``separate" phase will appear in a version of Fig.~\ref{fig:phase-1d-lambda} constructed with a 3D solver. {This suggestion is further bolstered by the parallel with atomic two-field BECs in traps, where the symmetry of the trap (in our case, spherical symmetry) can be broken by the ground state of the two species. \cite{2000JPhB...33.4017T}} However, showing this symmetry breaking in our case is numerically challenging. The total energy is very sensitive to numerical resolution, which is necessarily smaller in the 3D simulations than their 1D counterparts. Part of the problem lies in the competing numerical needs of each type of energy as the system transitions into the separate phase. As the two solitons of each field begin to separate, the numerical system requires very fine spatial resolution $\Delta x$ to classify when the two centres of mass cease to become co-incident at some minimum distance $d = \Delta x$. As their separation grows so does the side of the numerical box $L$ , both to accommodate $d$ and to minimize the numerical effects of the boundary of the box. However, the pressure to keep $\Delta x$ small is not lessened due to the quantum energy, which is sensitive to the gradients of each soliton's profile. Consequently, in the parts of parameter space where we expect the hollow and separate phases to dominate the grid size required $n_g = \left( L/\Delta x\right)$ is very large, putting both a memory and integration time strain on the numerical calculations. 

While full 3D simulations are beyond the scope of this work, we produced some preliminary 3D numerical results that support our conclusions presented above. We produced an alternate version of our phase diagram (Fig.~\ref{fig:phase-1d-lambda}) by comparing the energy of 1D solutions from \texttt{nSPIRal} placed on a 3D grid and the energy of two solitons placed with their centres of mass at a range of distances. The points where the separate solitons minimized the total energy for any distance $d$ greater than a single grid point was classified as ``separate", and the minimizing distance $d$ recorded. The resulting diagram maintains the hollow phase for a much reduced range of parameters and prefers the separate phase as $\lambda_{ij}$ grows. The same results also suggest the distance $d$ which minimizes the total energy similarly favors $d = 0$ on when $\lambda_{ij}\rightarrow 0$ and a growing $d$ as $\lambda_{ij}\rightarrow \infty$, with some modulation from the $m_2/m_1$ and $M_2/M_1$ choice. However, the results also display features that may be the result of numerical artifacts, such as ``islands" of hollow phases within the expected separate space. We expect these are due to numerical issues (discussed in the next section), but thorough verification of the results grew beyond the scope of this work. For this reason, careful construction of a full 3D phase diagram is deferred to work in our near future.

\begin{figure}
    \centering
    \includegraphics[width = 0.8\linewidth]{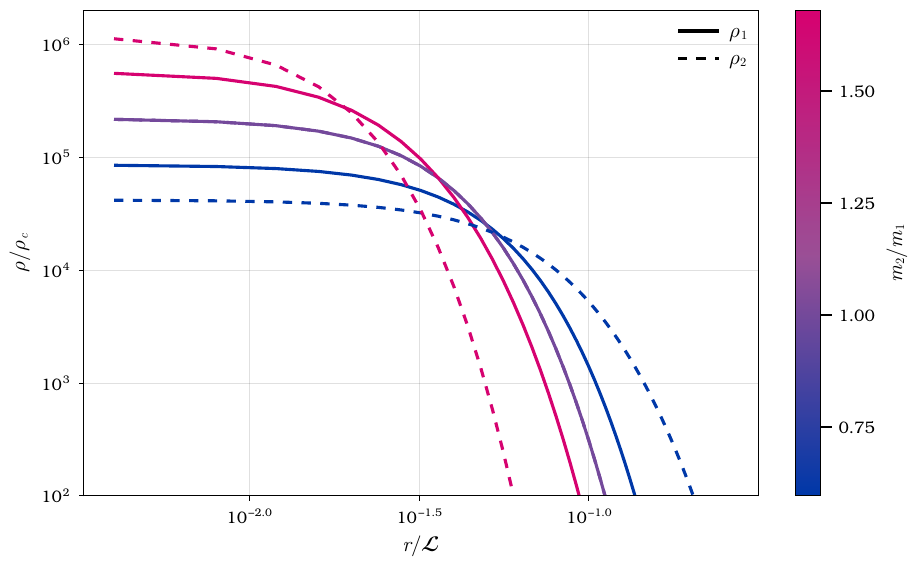}
    \caption{%
        Density profiles of nested solitons, computed by imaginary time relaxation in \texttt{nSPIRal}
        The solid lines denote $\rho_1$ and the dashed lines denote $\rho_2$.
        The color scale denotes the ratio $m_2/m_1$ of the particle masses.
        In each case the total masses are $M_1 = M_2 = 50 \mathcal{M}$ and $\Lambda = 0$.
    }
    \label{fig:profile_nested}
\end{figure} 

\section{Discussion}

This work uses 1D simulations to show that in scenarios with two gravitating scalar fields with repulsive interspecies interactions, there are at least three distinct phases of ground states with different phenomenological implications. This includes two nested or miscible phases, where the two fields' centres of mass are coincident, and one separate or immiscible phase, where the overall system breaks spherical symmetry. Work is ongoing to calculate ground states without the assumption of spherical symmetry and use 3D simulations to further explore these scenarios. These calculations will allow a phase diagram to be constructed that includes the immiscible phase. 

Our results indicate that relaxing the assumption of spherical symmetry results in the two fields separating into axially symmetric states; with three or more fields, even axial symmetry will be broken in the immiscible phase. In general, the structure of ground state configurations for $n$ fields with self- and interspecies interactions can become very complicated. Naturally, this means the richness and sheer size of the available axiverse parameter space will limit the reach of our approach. But work to understand the phase diagram in cases with low species number $n$—the parameter space within our reach—is realistic and timely, as the range of possible halo profiles in the axiverse could be confronted with data from current observational efforts constraining CDM. For example, the ongoing search for dark matter halos below $10^6 \, M_{\odot}$ is meant to test CDM's prediction that halos are self-similar at all scales, but could be equivalently leveraged to impose limits on axiverse particle numbers and masses given a range of possible interspecies interaction strengths and, therefore, halo shapes. Furthermore, independent modeling suggests that nonlinear resonance can arise between two axion species of similar masses and equilibrate their abundances, thus making the type of scenario we examine here more likely \cite{2022PhRvD.106h3503C}.

Thus, in light of the increasing diversity of equilibria in scalar dark matter, we propose that ``soliton'' is no longer a universally descriptive term for the cores predicted to exist in ULDM models.
Not only can each field's ground state differ from the solitonic profile approximated by Eq.~\ref{eq:soliton_profile} in terms of appropriate slope and radius $r_s$, one field in the nested hollow phase ceases to resemble a core of any kind. At the same time, the overall density of the system most resembles a core in the nested phase, as illustrated by the middle panel in Fig.~\ref{fig:gs-types}. Indeed, in cases such as the one considered here where the profile and shape of each wavefunction can differ significantly from the overall density, it becomes especially important to be precise which of the two quantities terms such as ``soliton" or ``core" refer to.    

A similar disambiguation could be worthwhile when going further to consider structure formation in such a cosmology between the condensation of solitons and the formation of halo cores. 
One the topic of the former, Jain et al.\ \cite{Jain:2023ojg} found that solitonic cores condense much more slowly when there are multiple fields—the condensation of one field serving to ``frustrate" the efforts of the other—and depends keenly on the relative number densities and, in some cases, particle mass ratio. Their results match overlapping cases from Chen et al.\ \cite{2023PhRvD.108h3021C}, who considered condensation in vector dark matter. This might lead us to expect that core formation in halos proceeds more slowly in multifield scenarios due to the increased condensation time. However, Luu et al. \cite{Luu:2023dmi} investigated non-linear structure formation
in two-field cosmological simulations and found that it can proceed \textit{faster} than in the one-field case when driven by the higher small-scale power in the two-field scenario. Contrasting these seemingly opposite findings yields a few guidelines for the future: firstly, that the different parameter regimes can provide different qualitative behavior (as is also argued in this paper) and, secondly, that future work may benefit from delineating soliton condensation from core formation when making predictions about cosmological structure formation. For either process, non-gravitational interspecies interactions could either speed up or slow down timescales, with either having an effect on cosmological structure formation in the axiverse or other ultralight dark sectors. Sufficient impact would be testable with current and upcoming observational missions. 

There are several possible directions for future work. The most obvious next step is a better understanding of the exact nature of the ground states in the separate or immiscible phase. While we have shown that this phase exists and are confident that its ground states would exhibit axial symmetry, we have not calculated the precise solutions nor presented a full phase diagram. It is worth emphasizing that, while we have shown that there are parameters for which the spherically symmetric ground state is hollow and verified this in 3D \textit{for our range of solutions}, we have not shown that they are the configurations with \emph{lowest possible} energy. {Based on our preliminary 3D tests we do expect the hollow ground states to persist as solutions, albeit for a much more narrow parameter space.} Achieving the numerical resolution necessary to extend our 1D imaginary time solver onto a fully realized 3D grid would be a reasonable next step towards this goal. An ideal scenario would involve comparison with completely different methods of finding the ground state (or even full spectrum) of our system; we are investigating some alternative approaches. 
Furthermore, models with three or more species would admit even more complex ground states. Under the assumption of spherical symmetry, there would be regimes that admit fully nested ground states, and others in which the ground state is a mixture of nested and hollow. It is also  unclear how the strength of interspecies interactions required to enter an immiscible phase scales with the number of fields. For any number of species, self-interactions would additionally affect the boundaries between the miscible and immiscible phases. 

Moving beyond purely stationary ground states, it is important to be aware of the gap between a mathematical ground state of the system (solitonic or otherwise) and the shape and evolution of the centre of a numerically idealized halo, let alone a cosmological one. Important effects have been demonstrated in the literature, including the soliton's random walk motion within a halo (see e.g. \cite{Li_random_walk}) and its changing shape in different backgrounds mimicking astrophysical phenomena~\cite{Guo:2020, Blum:2021oxj}. Multifield ground states embedded in halos will certainly exhibit similar modulation, whether through changing interference patters in the halo (as in e.g. \cite{2023PhRvD.107h3014G, Glennon:2023jsp}), dependence of core stability on species ratios in individual halos (\cite{Huang_2023, Jain:2023ojg, 2024PhRvD.109d3507L}), or other effects yet to be explored. We are particularly interested in the behavior of our nested hollow and separate phases inside halos, and plan to investigate it further. Indeed, we emphasize that this is a necessary step before confronting our stationary results with observational data collected from realistic and dynamic systems. 

Finally, while the bulk of this work and its extensions considered above can be considered a mathematical phase-space exploration, the present authors' interest in the results stems mainly from the possible astrophysical implications and links between theories of quantum gravity such as string theory, their resultant dark sectors, and observational astronomy. Thus, longer term prospects for using our results include synthesizing limits on axiverse scenarios from string theory calculations and observational constraints, as well as mapping to predictions of other soliton-forming dark sectors to unify and advance different particle theories with similar cosmological consequences.

\section*{Acknowledgements}%
\label{sec:acknowledgements}

We thank Noah Glennon for discussions in the beginning stages of this project and Hoang Nhan
Luu for useful discussions of the manuscript. LZ would like to thank Nikhil Padmanabhan for help setting up the code that would eventually evolve to become $\mathtt{nSPIRal}$. 
We thank the administrative and facilities staff at the University of New Hampshire including Katie Makem-Boucher and Michelle Mancini. {The authors thank the anonymous referee for helpful comments that improved this article.}

Computations were performed on Marvin, a Cray CS500 supercomputer at UNH supported by the NSF MRI program under grant AGS-1919310.
AEM's contributions to this project were supported by DOE Grant DE-SC0020220. CPW's contributions to this project were supported by the Research Corporation of America's Cottrell Scholars program.
MN acknowledges NASA ROSES grants 80NSSC24K1489 and 24-ADAP24-0074 and contract number 80NM0018F0610 via a JPL sub-award.
This work was performed in part at Aspen Center for Physics, which is supported by National Science Foundation under Grant No. PHY-1607611. Research at Perimeter Institute is supported in part by the Government of Canada through the Department of Innovation, Science and Economic Development Canada and by the Province of Ontario through the Ministry of Colleges and Universities.

\appendix

\section{Cosmology \& Code Units}\label{app:code-units}

When performing numerical simulations, we use units defined according to a reference particle mass $m_0 = 10^{-22}$ eV. This sets the code units as follows:
\begin{align}
    \mathcal{L}
    &=
    {\left(\frac{8\pi \hbar^2}{3m_0^2 H_0^2 \Omega_{m0}}\right)}^{\frac{1}{4}}
    \approx
    121 {\left(\frac{10^{-23} \mathrm{eV}}{m_0}\right)}^{\frac{1}{2}} \;\mathrm{kpc},
    \\
    \mathcal{T}
    &=
    {\left(\frac{8\pi}{3H_0^2\Omega_{m0}}\right)}^{\frac{1}{2}}
    \approx 75.5 \;\mathrm{Gyr},
    \\
    \mathcal{M}
    &=
    \frac{1}{G} {\left(\frac{8\pi}{3H_0^2\Omega_{m0}}\right)}^{-\frac{1}{4}} {\left(\frac{\hbar}{m_0}\right)}^{\frac{3}{2}}
    \approx 7\times10^7 {\left(\frac{10^{-23} \mathrm{eV}}{m_0}\right)}^{\frac{3}{2}}M_\odot
\end{align}
Note that the particle mass $m_0$ is different from the mass in the fields, $\mathcal{M}$. Interaction strengths are measured in units of
\begin{equation}
    \Lambda_{jk}
    =
    \frac{\hbar^2}{2 m_0^3 G \mathcal{T}}
    \lambda_{jk}
    \mperiod
\end{equation}
We assume a Hubble parameter $H_0 = 70 \;\mathrm{~km/s/Mpc}$ and matter density $\Omega_{m,0} = 0.3$.
The density unit $\mathcal{M}/\mathcal{L}^3 = \rho_c$ is the critical density of the universe, and energy is measured in units of $\mathcal{M} \mathcal{L}^2 \mathcal{T}^{-2}$.

\section{Imaginary Time Solver}\label{app:numerics}

We find nested ground states by solving the spherically symmetric versions of the GPP equations of motion in imaginary time and using the code units described above. The steps in our imaginary time evolution obey a ``kick-drift-kick" algorithm (as in \cite{2018JCAP...10..027E, 2021PhRvD.104h3532G, chplUltraPaper, Musoke:2024umh}).
Schematically, this can be represented for $n$ iterations of timestep $h$ for wavefunction $i$ as:
\begin{align}
    \psi_i(t+n h)
    =
    \exp \left[-\frac{h}{2} V_{\rm{eff}}\right]
    \times \left(\prod^{n} \exp [h V_{\rm{eff}}] \exp \left[-\frac{h}{2} \nabla^2\right]\right) 
    \times \exp \left[\frac{h}{2} V_{\rm{eff}}\right] \psi_i(t) 
    \,,
\end{align}
where
\begin{equation}
    V_{\rm{eff}} = m_i \Phi + \frac{1}{4} \sum_j m_j \lambda_{i j}\left|\psi_j\right|^2 \,.
\end{equation}
What the equations above don't capture is that the gravitational potential $\Phi$ also gets updated at every timestep. Effectively, this is equivalent to $n$ iterations of: 
\begin{enumerate}
    \item evolving $\psi_i$ for a half-step $h/2$ in the gravitational potential $\Phi\left(t - \frac{h}{2}\right)$,

    \item  applying the kinetic operator to $\psi$, then re-calculating $\Phi$ from $\psi\left(t + \frac{h}{2}\right)$,

    \item evolving $\psi_i$ for another half-step given the new $\Phi$, 

    \item re-normalizing $\psi_i$ so that the total mass in each field is conserved.
\end{enumerate}

Additionally, \texttt{nSPIRal} introduces a hard boundary at a given maximum radial distance $r=r_{\text{max}}$. The boundary value of each field is fixed such that  $\left. r\psi_j \right|_{r = 0} = \left. r\psi_j \right|_{r_{\text{max}}} = 0$; the latter is equivalent to specifying $\psi_j(r_{\text{max}}) = 0$. We therefore take care to use only large enough $r_{\rm{max}}$ such that $\psi_j(r_{\rm{max}}) \rightarrow 0$ for each field. 

In addition to $r_{\rm{max}}$, we must specify our grid size $n_g$, number of timesteps $n$, and initial conditions for the $\psi_j$ profiles. We have found that a large swatch of our parameter space is relatively insensitive to changes in these three parameters—except in vicinity of the degenerate case  $m_1 = m_2$, $M_1 = M_2$, and  $\psi_1(r) = \psi_2(r)$, as discussed in Sec.~\ref{sec:numerical-results}. Though not central to our results in this paper, we have also produced ground state profiles assuming $\lambda = 0$ and plotted it in Fig.~\ref{fig:profile_nested} in order to verify that the behavior matches predictions in the literature.

\bibliographystyle{JHEP}
\bibliography{bibliography}

\end{document}